\documentclass[conference,10pt]{IEEEtran}
\IEEEoverridecommandlockouts

\usepackage{amsmath,amssymb,amsfonts}

\usepackage{graphicx}
\usepackage{textcomp}
\usepackage{xcolor}
\def\BibTeX{{\rm B\kern-.05em{\sc i\kern-.025em b}\kern-.08em
    T\kern-.1667em\lower.7ex\hbox{E}\kern-.125emX}}

\usepackage{enumitem}
\usepackage{environ}
\usepackage{tabularx}
\usepackage{soul}

\usepackage{hyperref}
\hypersetup{
    colorlinks=true,
    linkcolor=black,
    urlcolor=black,
    citecolor=black
}

\usepackage{cleveref}
\crefformat{chapter}{\S#2#1#3}
\crefformat{section}{\S#2#1#3}
\crefformat{subsection}{\S#2#1#3}
\crefformat{subsubsection}{\S#2#1#3}

\usepackage{xspace}
\usepackage{subfig}
\usepackage{booktabs}
\usepackage{multirow}
\usepackage{fancyvrb}

\usepackage{balance}
\usepackage[noadjust]{cite}
\usepackage{float}

\definecolor{navy}{HTML}{2F729C}
\definecolor{darkgray}{HTML}{323232}
\definecolor{seagreen}{HTML}{228B22}

\newcommand{\baseline}{\textbf{\emph{baseline}}\xspace}
\newcommand{\custom}{\textbf{\emph{custom}}\xspace}
\newcommand{\optimizer}{\textbf{\emph{optimizer}}\xspace}

\renewcommand{\paragraph}[1]{\vspace{2.5pt}\noindent\textbf{#1.}}

\usepackage{amsmath, amsfonts, mathtools, amsthm} 

\usepackage[framemethod=TikZ]{mdframed}
\usepackage{thmtools}
\mdfsetup{skipabove=1em,skipbelow=0em, innertopmargin=5pt, innerbottommargin=6pt}
\definecolor{DimGray}{rgb}{0.41, 0.41, 0.41}
\colorlet{thmgraycolor}{DimGray!70!black}

\makeatletter
\declaretheoremstyle[
	headfont=\bfseries\color{thmgraycolor},
	bodyfont=\normalfont\itshape,
	mdframed={
			innertopmargin=2pt,
			innerleftmargin=8pt,
			innerrightmargin=8pt,
			innerbottommargin=5pt,
			linewidth=2pt,
			rightline=false, topline=false, bottomline=false,
			linecolor=black, backgroundcolor=DimGray!5,
		}
]{thmgraybox}
\makeatother

\declaretheorem[style=thmgraybox, numbered=yes, name=Observation]{observation}

\begin{document}
\title{Idiosyncrasies of Programmable Caching Engines}

\author{{\bf Workshop Paper}\vspace{0.2em}\\
	\begin{minipage}{0.7\textwidth}
		\centering
		José~Peixoto, Alexis~Gonzalez\textsuperscript{*}, Janki~Bhimani\textsuperscript{*}, Raju~Rangaswami\textsuperscript{*}, Cláudia~Brito, João~Paulo, Ricardo~Macedo\vspace{0.3em}\\
		\textit{INESC TEC \& University of Minho \hfill \textsuperscript{*}Florida International University}
	\end{minipage}
}

\maketitle

\begin{abstract}
	Programmable caching engines like CacheLib are widely used in production systems to support diverse workloads in multi-tenant environments.
	CacheLib's design focuses on performance, portability, and configurability, allowing applications to inherit caching improvements with minimal implementation effort.
	However, its behavior under dynamic and evolving workloads remains largely unexplored.
	This paper presents an empirical study of CacheLib with multi-tenant settings under dynamic and volatile environments.
	Our evaluation across multiple CacheLib configurations reveals several limitations that hinder its effectiveness under such environments, including rigid configurations, limited runtime adaptability, lack of quality-of-service support and coordination, which lead to suboptimal performance, inefficient memory usage, and tenant starvation.
	Based on these findings, we outline future research directions to improve the adaptability, fairness, and programmability of future caching engines.
\end{abstract}


\section{Introduction}
\label{sec:introduction}

Data-intensive systems, such as databases, key-value stores, content delivery networks (CDNs), and machine learning engines, have become a fundamental part of modern I/O infrastructures \cite{fengScaling2021}.
To process large volumes of data with high throughput and low latency, these systems rely on in-memory caching.
Typically, each system employs a cache optimized for its specific requirements, including read-write ratio, access distribution, I/O granularity (block, file, object), and concurrency model (e.g., single vs. multi-tenant) \cite{bergCacheLib2020, stefanoviciSoftwaredefined2015}.

Despite their differences, caching systems often share common goals and face similar challenges (\textit{e.g.}, interfaces, eviction policies, memory allocation strategies).
The lack of a unified cache abstraction leads to fragmented features, duplicated engineering efforts, and significant maintenance overhead.
To address these limitations, Meta introduced CacheLib~\cite{bergCacheLib2020}, a general-purpose, programmable caching engine that provides a common set of building blocks for designing high-performance caches.
Through a flexible and extensible API covering cache indexing, thread-safe structures, eviction policies, memory management, and multi-tenancy isolation, CacheLib enables the design of caches fine-tuned for diverse data-intensive systems.
Its success stems from configurability and performance portability, allowing applications to inherit caching improvements without major reimplementation.

Today, CacheLib powers more than 70 Meta services (\textit{e.g.}, CDNs, key-value stores, recommendation engines, databases)~\cite{bergCacheLib2020}, Pelikan.io, and several research initiatives \cite{noauthorCacheLib2021}.
For ease of management and cost efficiency, multiple services and applications are often co-located on the same compute node, each managing its own cache instance.
Even a single application may employ multiple caches (\emph{e.g.}, with different eviction policies and sizes) to serve distinct workloads~\cite{bergCacheLib2020,bergerRobinHood2018}.

While CacheLib provides a strong foundation, important questions remain in the context of shared, multi-tenant deployments: \textit{i)} how does CacheLib simplify memory management under multiple tenants? \textit{ii)} how does it handle workloads with evolving access patterns and dynamic resource demands?

This paper presents the first empirical study of CacheLib's performance and adaptability in dynamic, multi-tenant environments.
Our study reveals several limitations that open new research directions. 

\paragraph{Rigid configuration model}
CacheLib's configuration model is too rigid for the dynamic nature of co-located workloads.
Although it exposes many tuning knobs (\emph{i.e.,} eviction policies, memory allocation, and rebalancing schemes), these parameters are typically set at initialization and do not adapt to changing workloads (\emph{e.g.,} shifting access distributions, tenants joining or leaving), leading to suboptimal performance.

\paragraph{Limited adaptability}
CacheLib includes mechanisms such as the \emph{pool optimizer} to adjust memory allocations at runtime.
However, our study shows these mechanisms are often ineffective under shifting workloads, resulting in performance variability and inefficient memory use.

\paragraph{Lack of quality-of-service (QoS) support}
CacheLib's aggressive reallocation strategies prioritize global cache efficiency over individual tenants.
Without QoS guarantees, prioritization, or differentiation mechanisms, this behavior can lead to starvation, especially for latency-sensitive workloads that share resources with more aggressive tenants.

\paragraph{Lack of inter-instance coordination}
Multiple CacheLib instances can run on the same compute node but operate in isolation.
This lack of coordination prevents efficient resource utilization and holistic memory rebalancing, especially under skewed workloads where some instances are underprovisioned while others are overprovisioned.

In summary, our findings show that while CacheLib offers key building blocks for multi-tenant environments, it falls short in handling the complexities introduced by dynamic workloads and shared resources.
To bridge this gap, we identify several research directions to improve CacheLib's adaptability, inter-instance coordination, and QoS support, making it more effective and robust for modern data-intensive applications.

\section{Background}
\label{sec:background}

This section provides background on the CacheLib engine, outlining its design, core functionalities, and applicability.

\begin{figure}[t]
	\centering
	\includegraphics[width=1\linewidth,keepaspectratio]{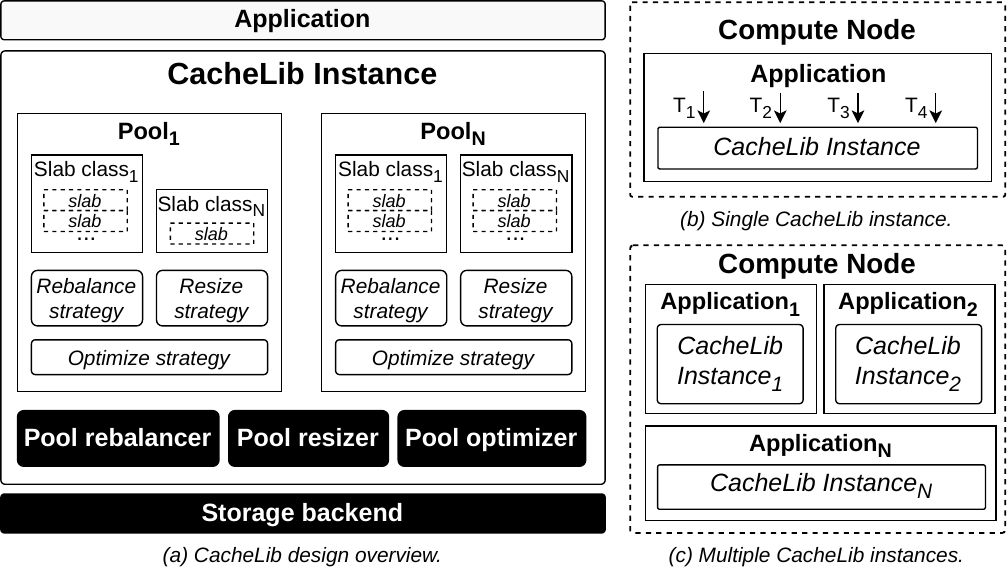}
	\caption{CacheLib design overview (a) and typical applicability models in production (b-c).}
	\label{fig:cachelib-overview}%
	\vspace*{-10pt}
\end{figure}

\subsection{CacheLib overview}
\label{subsec:overview}

CacheLib~\cite{bergCacheLib2020} is an embeddable caching library developed by Meta, designed for high performance in concurrent environments.
Its architecture emphasizes memory efficiency, workload isolation, and runtime adaptability.
It exposes a simple, thread-safe interface where users can store and retrieve items using a key-value pair abstraction.
At its core, CacheLib comprises four main components, as depicted in Fig.~\ref{fig:cachelib-overview}~\emph{(a)}.
The \emph{memory pool}, an isolated region with independent allocation strategies, eviction policies, and memory limits, stores and serves data items.
The \emph{pool rebalancer} and \emph{pool resizer} minimize memory waste and fragmentation by moving memory within and across pools.
The \emph{pool optimizer} aims at improving overall memory usage by dynamically resizing each pool.

\paragraph{Memory allocation and fragmentation management}
CacheLib employs a slab-based memory allocation scheme to minimize external memory fragmentation.
Rather than allocating variable-sized chunks, it partitions memory into fixed-size slabs, each associated with a (slab) class that stores items of similar size.
Each slab class manages its own memory region, and a single memory pool may contain multiple slab classes.
Once memory has been allocated across slab classes, the system enters a static state, which can lead to \emph{slab calcification}.
This phenomenon occurs when slabs are assigned to item sizes that are no longer accessed, leading to wasted memory and allocation failures for other slab classes.
To mitigate this, CacheLib integrates a \emph{pool rebalancer}, a background worker that \emph{1)} identifies imbalanced slab classes within a memory pool, \emph{2)} selects a victim class with surplus slabs, and \emph{3)} transfers slabs to a receiver class with higher demand.
Moreover, each pool employs a configurable \emph{rebalance strategy} to guide the slab distribution, such as minimizing allocation failures or maximizing per-item hit ratio.

\paragraph{Multi-tenancy support}
To support multiple concurrent workloads, CacheLib enables the creation of numerous memory pools, each isolating a specific workload or tenant, enabling flexible memory partitioning within the same CacheLib instance.
By default, each pool is assigned a static memory limit.
While users can adjust this limit at runtime, increasing a pool's allocation does not guarantee it will immediately receive more memory, as slabs may remain held by other pools until explicitly rebalanced.
To enforce these limits, CacheLib implements a \emph{pool resizer}, a background worker that \emph{1)} identifies pools exceeding their configured memory limits, \emph{2)} selects slabs from pools exceeding their capacity, and \emph{3)} transfers them to underprovisioned ones.
This process allows the system to better enforce memory limits, but does require manual intervention to initiate resizing.
Moreover, each memory pool implements a \emph{resize strategy} (\emph{e.g.,} maximize hit ratio, minimize allocation failures) to guide the resize process.

\paragraph{Adaptive memory management}
To reduce manual configuration overhead, CacheLib implements the \emph{pool optimizer}, a background worker for automatic pool resizing~\cite{noauthorCacheLib2021}.
It continuously monitors the workload, analyzes access patterns at the tail of the eviction queue to estimate workload pressure, and proposes new memory limits for each pool to enhance overall hit ratio.
To enforce these limits, the pool optimizer and resizer must operate together.

\subsection{CacheLib request workflow}
\label{subsec:workflow}

The request workflow in CacheLib is divided into two parts: the \emph{foreground} flow, which handles read and write requests submitted by the application, and the \emph{background flow}, which manages internal system operations.

\paragraph{Foreground flow}
Applications interact with the cache by inserting and retrieving data objects through a key-value pair abstraction. 
On writes, the application first attempts to allocate memory for the object within the target pool. If the allocation succeeds, the application copies the data into the allocated region, and only then can the transaction be committed to the cache. It is the application's responsibility to ensure data persists in durable storage.
On reads, given a key, CacheLib searches across all memory pools and returns the object if found, or if a cache miss occurs, otherwise, leaving the responsibility of retrieving data from persistent storage to the application.

\paragraph{Background flow}
The background flow is responsible for maintaining the cache's internal structure and adapting to workload changes through background workers, including the \emph{pool rebalancer}, \emph{resizer}, and \emph{optimizer} (\cref{subsec:overview}).
These operations are executed periodically, asynchronously, and independently of the foreground flow, allowing CacheLib to adapt to workload changes without interrupting application service.

\subsection{Usage in production environments}
\label{subsec:production}

CacheLib supports a range of deployment scenarios, depending on the number of applications and tenants involved.
This section highlights two common production setups, which serve as the foundation for the discussion in the following sections.

\paragraph{Single instance, multiple tenants}
As depicted in Fig.~\ref{fig:cachelib-overview}~\emph{(b)}, a single application or service, such as CDNs, key-value stores, social-graph systems, and databases, can use just a single CacheLib instance to support multiple tenants~\cite{bergCacheLib2020}.
Tenants are typically represented by distinct threads, processes, or components with different request patterns and memory needs.
Each tenant is assigned a dedicated memory pool, ensuring isolation and predictable performance.

\paragraph{Multiple instances}
As depicted in Fig.~\ref{fig:cachelib-overview}~\emph{(c)}, multiple CacheLib instances are deployed on the same compute node, each serving a different application or service, which is common in complex data-intensive software stacks, such as those used by Uber, Amazon, and more~\cite{UberInfrastructure:2021:Fu,AmazonRedshift:2022:Armenatzoglou,OpenConnect:2018:Bottger}.
Instances operate in isolation, with individual configurations and allocation policies, configured with a fixed memory limit.
\section{Understanding CacheLib Performance}
\label{sec:motivation}

\begin{figure*}[t]
	\centering
	\includegraphics[width=\textwidth]{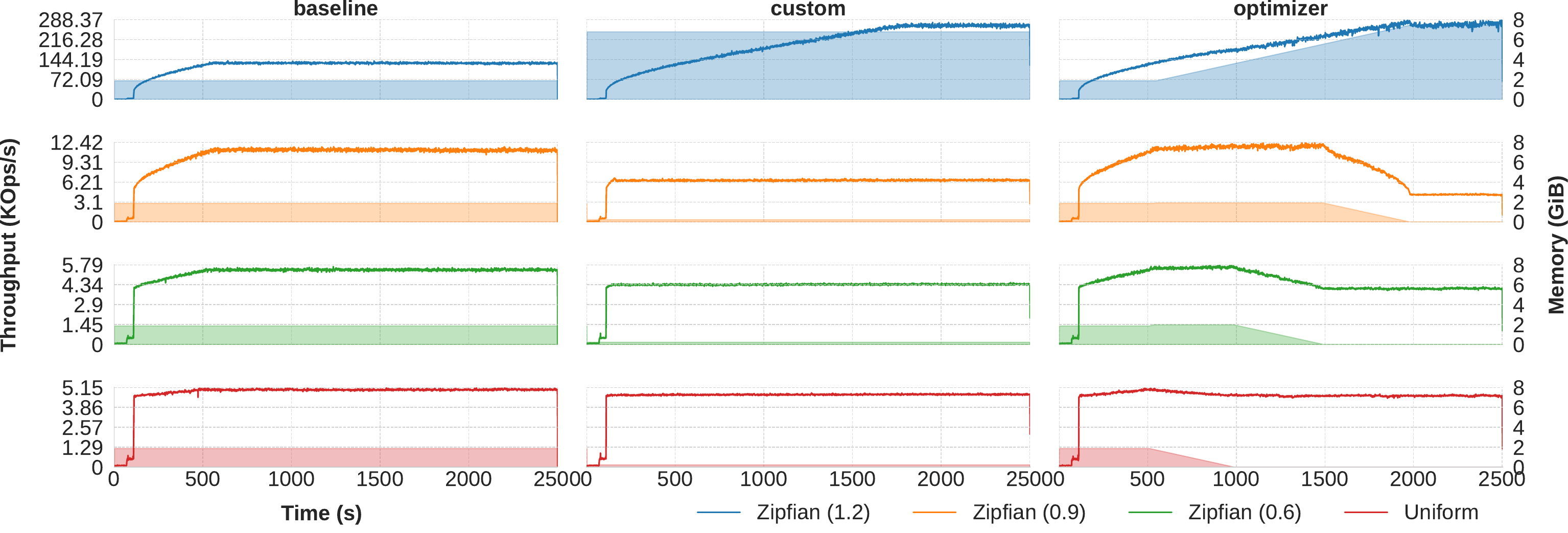}
	\caption{\textbf{Per-tenant performance comparison across CacheLib {baseline}, {custom}, and {optimizer} setups (columns).} 
	Each row depicts the throughput (kops/s) and allocated memory (GiB) over time of each tenant (T$_1$ to T$_4$). 
	The left y-axis presents the throughput (line), while the right y-axis presents the memory allocated (filled area).}
	\label{fig:motivation-1-throughput-memory-threads}
	\vspace*{-12.5pt}
\end{figure*}

To understand the performance and adaptability of CacheLib under dynamic and heterogeneous environments, we seek to answer the following questions:
\begin{itemize}[leftmargin=*]
	\item How does CacheLib perform under varying workload request popularity distributions?
	\item How does CacheLib perform under different memory partitioning?
	\item Can CacheLib adapt to dynamic and evolving workloads?
\end{itemize}

\paragraph{Hardware and OS configurations}
Experiments were conducted on compute nodes of the Deucalion supercomputer equipped with 2$\times$ 64-core AMD EPYC 7742 processors, 256~GiB of memory, and a 480~GiB SSD, running RockyLinux 8.
Software-wise, we used \href{https://github.com/facebook/CacheLib/releases/tag/v20231101_RC1}{CacheLib v20231101}.

\paragraph{Methodology}
The experimental testbed consists of three components: a benchmark that acts as the \emph{application} and generates requests with varying access distributions, a CacheLib instance that caches the application's read requests, and RocksDB as the persistent storage backend.
Write operations are always submitted to RocksDB, and to avoid cache incoherence, the write operation is propagated to the cache if the item was previously cached.
For reads, requests are first submitted to CacheLib -- on a cache hit, it returns the corresponding item to the application; on a cache miss, the application fetches the item from persistent storage and inserts it in the corresponding CacheLib memory pool.
Further, to isolate the performance benefits of CacheLib, both RocksDB's internal block cache and the page cache were disabled.

As in previous studies (\cite{shuePerformance2012,yangSegcache2021,asadAdaptCache2016}), we evaluated CacheLib's performance in a heterogeneous multi-tenant environment by conducting experiments with 4 tenants, each configured with different workload characteristics and memory limits.

\paragraph{Workloads}
Experiments were conducted using read-only workloads, as it is common in evaluating caching systems~\cite{bergCacheLib2020,qiuFrozenHot2023,shuePerformance2012}.
Each tenant is configured with a distinct workload distribution, emulating access patterns observed in production environments~\cite{yanglarge2020}.
Specifically, tenants T$_1$, T$_2$, and T$_3$ were assigned with \emph{Zipfian} distributions with skew factors of 1.2, 0.9, and 0.6, respectively, while tenant T$_4$ was configured with a \emph{uniform} distribution.
Further, each tenant operated on an exclusive key-space containing 20 million unique items, each 1~KiB in size, totaling approximately 20~GiB of data.
Unless stated otherwise, all tenants execute their workloads concurrently over 2500 seconds.
Additionally, before workload execution, the 20 million unique key-value pairs of 1~KiB each are pre-loaded into the storage backend for each tenant, totaling approximately 80~GiB of data. As such, the size of the CacheLib instance was configured to store only up to 10\% of the dataset, similarly to previous studies (\cite{qiuFrozenHot2023,wuzExpander2016}), resulting in a total memory capacity of 8~GiB partitioned into four memory pools, each dedicated to a specific tenant.

\paragraph{Setups}
To explore the impact of different memory management strategies, with three CacheLib configurations:
\emph{\textbf{baseline}} refers to the default CacheLib, where only the \textit{pool rebalancer} worker is enabled; \emph{\textbf{custom}} extends the baseline setting by integrating a custom memory allocation policy which, in combination with the \emph{pool resizer}, enforces dynamic memory limits defined by the user for each pool; and \emph{\textbf{optimizer}} enables \emph{pool resizer} and \emph{pool optimizer} workers.
Workers from both \custom and \optimizer execute every second; the \emph{pool resizer} was set with a resize strategy to maximize hit ratio, which best suits our needs. 

\subsection{Memory partitioning under static workloads}
\label{subsec:eval-static}

We begin by evaluating the performance and adaptability of different CacheLib configurations in a static environment, where all tenants start and complete their workloads simultaneously. We use the following configurations that apply distinct memory allocation policies:
\begin{itemize}[leftmargin=*]
	\item \textbf{\emph{Baseline}:} memory is allocated uniformly across all tenants, with each pool receiving 25\% of the total cache capacity.
	\item \textbf{\emph{Custom}:} favors the tenant with the highest Zipfian skew, allocating 91\% of the cache capacity to T$_1$, while the remaining space is evenly distributed across the other tenants.
	\item \textbf{\emph{Optimizer}:} initially adopts a uniform allocation but dynamically adjusts the memory distribution during execution based on the observed cache performance metrics, and is tuned to maximize overall throughput.
\end{itemize}

\begin{figure}[ht!]
	\includegraphics[width=0.5\textwidth]{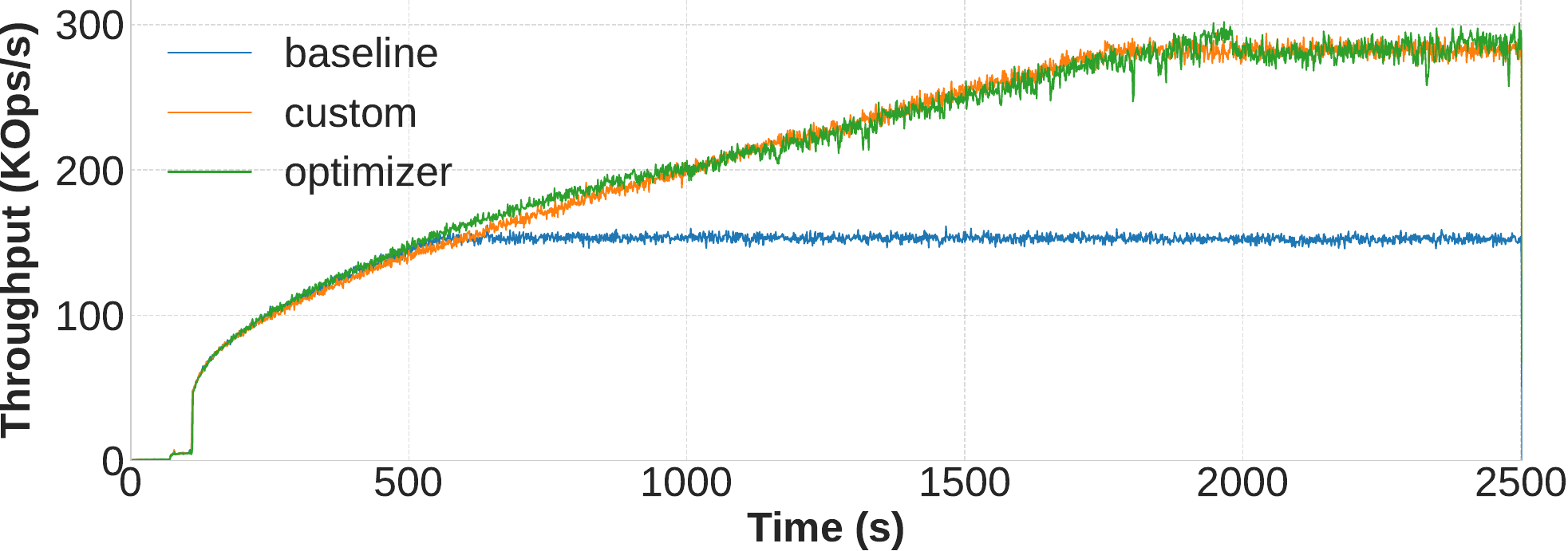}
	\caption{\textbf{Overall throughput under different allocation policies.} Workload-aware memory allocation substantially outperforms uniform policies.}
	\label{fig:motivation-1-throughput-global-comparison}
	\vspace*{-10pt}
\end{figure}

\paragraph{Results}
Fig.~\ref{fig:motivation-1-throughput-memory-threads} and \ref{fig:motivation-1-throughput-global-comparison} illustrate, respectively, the per-tenant and global throughput and allocated memory under the three CacheLib configurations.
As expected, throughput varies significantly across tenants due to their different data distributions.
In the \baseline setup, after the initial cache warm-up, T$_1$ reaches $\approx$140~kops/s, outperforming T$_2$, T$_3$, and T$_4$ by 14$\times$, 25$\times$, and 28$\times$, respectively.
These results highlight that uniform memory allocation can lead to imbalanced performance when tenants exhibit heterogeneous access patterns. 

The \custom configuration shows that favoring highly skewed workloads yields significant performance gains.
As depicted in Fig.~\ref{fig:motivation-1-throughput-global-comparison}, the overall system throughput improves up to 1.8$\times$ compared to the baseline, with T$_1$ experiencing a throughput of up to 280~kops/s.
This is because the number of I/O requests to the backend grows inversely with the hit ratio, meaning that small increases in hit ratio lead to significant throughput improvements when the hit ratio is already high.
In the case of T$_1$, which follows a highly skewed distribution, further gains in hit ratio/throughput required a disproportionately larger share of memory, thus the 91\% allocation.
However, if the goal was instead to maximize hit ratio across all tenants or to ensure fairness, this partitioning might not be optimal.

\begin{observation}
	Under heterogeneous workloads, uniform memory allocation can lead to severe performance imbalance.
	Fine-tuning cache partitioning according to workload distribution can significantly improve throughput.
\end{observation}

As depicted in Fig.~\ref{fig:motivation-1-throughput-memory-threads}, the \optimizer setup begins with the same uniform memory distribution as the \baseline but gradually reallocates memory based on observed workload behavior. 
This process, however, requires a warm-up period to collect enough statistics.
Only after $\approx$500 seconds, the \optimizer begins adjusting memory allocations effectively, converging toward an improved state.
Despite this initial delay, as observed by Fig.~\ref{fig:motivation-1-throughput-global-comparison}, the overall throughput of \optimizer is higher than the \baseline configuration, reaching $\approx$290~kops/s, a 1.85$\times$ improvement under the same memory capacity.

\begin{observation}
	Dynamically reallocating memory based on workload characteristics can outperform static configurations, even under identical resource constraints.
\end{observation}

Interestingly, a closer inspection (Fig.~\ref{fig:motivation-1-throughput-memory-threads}) reveals that the \optimizer reallocates nearly the entire cache capacity to T$_1$, increasing its memory pool from 2~GiB to $\approx$8~GiB.
This leads to starvation of the remaining tenants, whose throughput significantly drops, even falling below the performance observed under the \baseline setup.
While such behavior might be suited for environments where maximizing the global throughput is the primary goal, it is incompatible with scenarios that require per-tenant quality-of-service (QoS) guarantees or enforce different priority levels.
Under these scenarios, even if global throughput is improved, the \optimizer's aggressive reallocation strategy compromises the performance isolation expected between tenants.

\begin{observation}
	Without QoS or prioritization mechanisms, CacheLib's dynamic reallocation strategy can lead to starvation, compromising the performance isolation across tenants and resulting in unfair memory attribution.
\end{observation}

\begin{figure*}[t]
	\includegraphics[width=\textwidth]{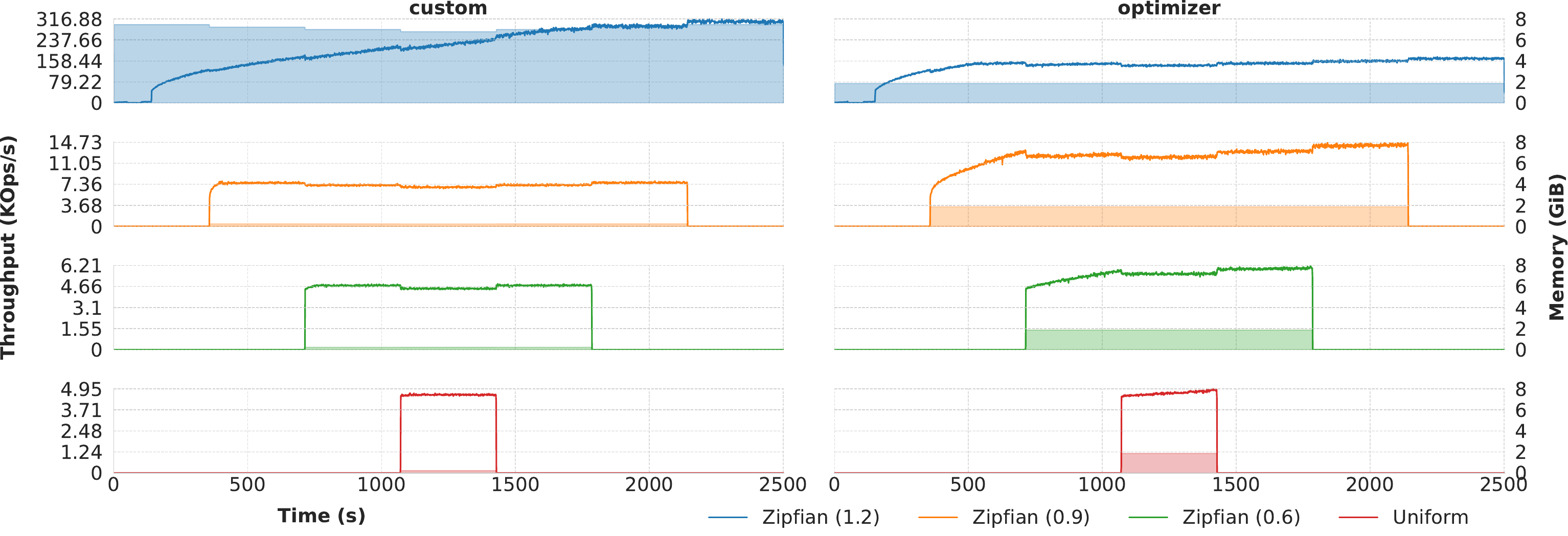}
	\caption{\textbf{Per-tenant performance comparison across CacheLib custom and optimizer setups under dynamic workloads.}
		Each row depicts the throughput (kops/s) and allocated memory (GiB) over time of each tenant (T$_1$ to T$_4$).
		The left y-axis presents the throughput (line), while the right y-axis presents the memory allocated (filled area).}
	\label{fig:motivation-2-throughput-memory-threads}
	\vspace*{-12.5pt}
\end{figure*}

\subsection{Memory partitioning under dynamic workloads}
\label{subsec:eval-dynamic}

Both \baseline and \custom configurations rely on static memory allocation, where the memory assigned to each tenant is fixed at setup time.
While this approach can be effective in stable scenarios, where workload characteristics and system settings remain constant over time, this may not hold for more volatile environments.

We now evaluate different CacheLib setups under such environments, where tenants enter and leave the system at different points in time.
Specifically, tenants are added to the system sequentially every 500~seconds, and after all tenants are active for 500 seconds, they leave the system in reverse order of arrival.
The initial memory allocations across tenants follow the same setup as in \cref{subsec:eval-static}.
For the \custom setup, the allocation policy was adapted with the following rules: \emph{1)} at the start up time, no tenant is allocated with more memory than its expected demand, and \emph{2)} any leftover memory capacity if available is proportionally distributed among active tenants to prevent underutilization of memory resources.
Finally, the \baseline configuration is omitted from this experiment, as it is not capable of adapting to dynamic workloads, thus would exhibit the same behavior as in \cref{subsec:eval-static}.

\begin{figure}[t]
	\includegraphics[width=0.5\textwidth]{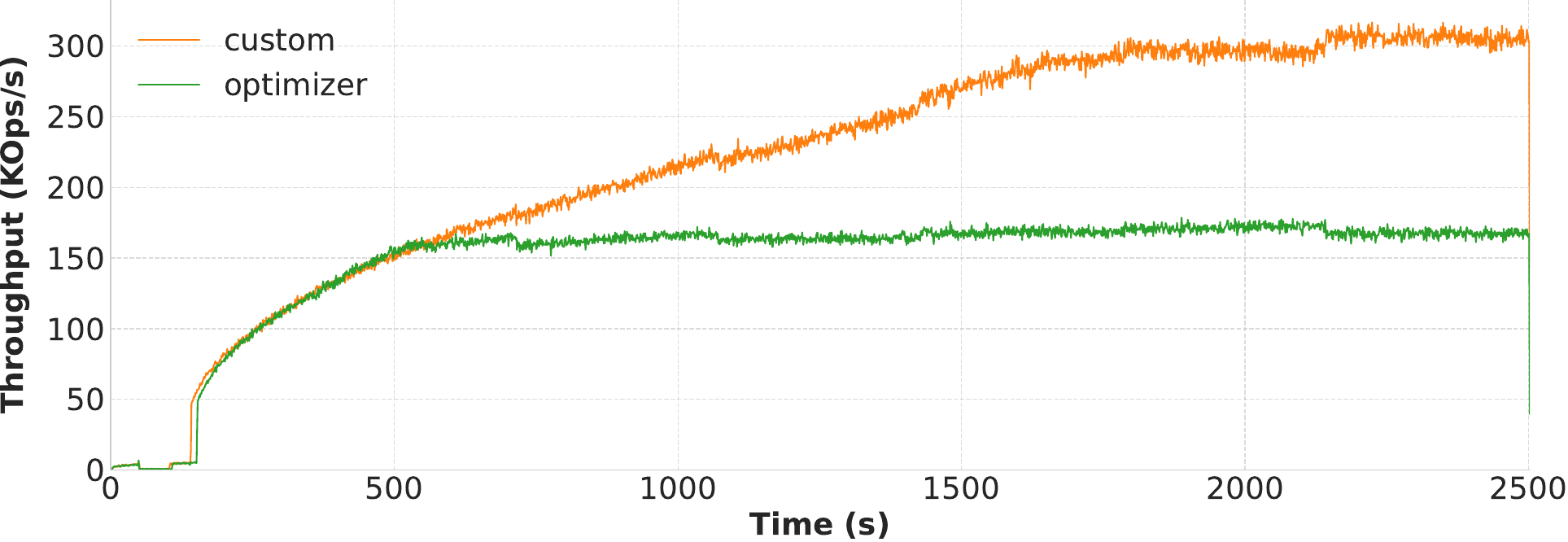}
	\caption{\textbf{Overall throughput under dynamic workloads.} Workload-aware reallocation outperforms static or delayed adaptation.}
	\label{fig:motivation-2-throughput-global-comparison}
	\vspace*{-10pt}
\end{figure}

\paragraph{Results}
Fig.~\ref{fig:motivation-2-throughput-memory-threads} depicts the throughput and memory allocation per tenant under the \custom and \optimizer setups.
Contrary to the previous results (\cref{subsec:eval-static}), the \optimizer fails to reallocate memory dynamically and converge to an improved state since tenants are active for too short a period, resulting in insufficient data to trigger reallocation. Nevertheless, even if tenant interleaving periods were larger, the \optimizer could still fail to converge due to erroneous memory allocation decisions under the rapidly changing nature of the workloads.
In any case, the \optimizer fails to adapt and remains with a static memory allocation of 25\% of the total cache capacity to each tenant, achieving a maximum throughput of $\approx$170~kops/s.

On the other hand, as depicted in Fig.~\ref{fig:motivation-2-throughput-global-comparison}, the \custom configuration achieved a peak throughput of $\approx$315~kops/s, a 1.85$\times$ improvement over the \optimizer under the same memory constraints.
This improvement stems from two key factors.
First, the startup memory allocations of \custom are better aligned with the expected workload characteristics, ensuring each tenant receives a memory share closer to its actual demand.
Second, the custom policy reclaims and redistributes unused memory proportionally among active tenants, improving overall resource usage and throughput.
For instance, during the first 500 seconds of the experiment, when only T$_1$ is active, while the \optimizer limits T$_1$ to 25\% of the cache, the \custom setup allocates the full cache capacity, maximizing its performance.
As additional tenants join, the custom policy continues to dynamically adjust memory allocations proportionally, ensuring effective cache usage.

\begin{observation}
	To maximize cache effectiveness in dynamic environments, it is fundamental to combine workload-aware memory allocation primitives with proportional redistribution of unused memory resources.
\end{observation}

\section{Discussion}
\label{sec:discussion}

CacheLib made significant contributions to the design of programmable caching engines, particularly for multi-tenant environments. Despite offering a \emph{pool optimizer} for better resource management, our study highlights that CacheLib still faces challenges in handling dynamic workloads on multi-tenant deployments, which opens up the path for new research.

\paragraph{Handling dynamic workloads}
Our study shows that the \textit{pool optimizer} cannot properly handle dynamic tenants, leading to inefficient memory allocation and underutilization of available resources.
In particular, when tenants frequently join or leave the system, such as in applications with background jobs (\emph{e.g.,} RocksDB compactions) or social networks with services experiencing idle periods followed by bursts of activity (\emph{e.g.,} Twitter~\cite{yanglarge2020}), the \emph{pool optimizer} fails to react promptly and adequately.
To address this, the prediction mechanism could be improved using techniques such as Miss-Ratio Curves (MRCs) to describe how hit-ratio, throughput, or latency varies with distinct pool sizes~\cite{kollerCentaur2015, waldspurgerEfficient2015}.

\paragraph{Handling multiple instances}
CacheLib instances operate in isolation, leading to the absence of global coordination.
Handling dynamic workloads across multiple instances becomes more challenging, as the pool optimizer only manages pools within the same instance, preventing memory from being reallocated across instances and limiting overall resource efficiency.
This isolation also hinders the ability to enforce QoS guarantees to prioritize critical tenants over less critical ones.
Solving this challenge will require a coordinated resource management layer across instances, with QoS-aware policies and, potentially, through a centralized controller capable of monitoring cache usage and enforcing tenant-specific performance targets across the entire infrastructure~\cite{macedoPADLL2023,mirandaCheferd2024}.

\paragraph{Related work and study generalizability}
This paper focuses on CacheLib, given its widespread adoption.
Nonetheless, several other caching solutions have been proposed to address multi-tenant workloads.
Dynacache~\cite{cidonDynacache2015}, Cliffhanger~\cite{cidonCliffhanger2016}, AdaptSize~\cite{bergerAdaptSize2017}, zExpander~\cite{wuzExpander2016}, FrozenHot~\cite{qiuFrozenHot2023}, LAMA~\cite{huLAMA2015}, AdaptCache~\cite{asadAdaptCache2016}, and SegCache~\cite{yangSegcache2021} focus on general cache optimizations, such as reducing latency, increasing throughput, improving hit ratios, or balancing memory usage, by employing different strategies, including workload monitoring, adaptive resizing, and probabilistic policies.
However, these approaches are tenant-oblivious, as they aim to maximize cache performance rather than providing tenant fairness and isolation, crucial for multi-tenant environments such as those supported by CacheLib.
This limitation often leads to aggressive tenants monopolizing cache resources, inadvertently evicting other tenants' working sets, leading to performance degradation and QoS violations~\cite{yanglarge2020,shuePerformance2012}.

As for systems that explicitly address QoS, Moirai~\cite{stefanoviciSoftwaredefined2015}, Pisces~\cite{shuePerformance2012}, Centaur~\cite{kollerCentaur2015}, and RobinHood~\cite{bergerRobinHood2018} implement tenant-aware management techniques to meet specific performance objectives. 
For instance, Centaur employs optimization algorithms, such as Simulated Annealing (SA) combined with Miss-Ratio Curves (MRCs), to optimize the cache partitioning and allocation of resources across multiple tenants to meet the specified performance constraints~\cite{kollerCentaur2015}. However, these systems are typically fine-tuned for specific workloads and storage scenarios, such as hypervisors and key-value stores, limiting their generalizability to broader multi-tenant caching environments like those targeted by CacheLib.
For example, a prediction model designed for block caches assumes uniform item sizes, whose assumption breaks down in key-value stores, which handle variable-sized items, leading to inaccurate performance estimations~\cite{kollerCentaur2015,waldspurgerEfficient2015}.
\section*{Acknowledgements}
This work was co-funded by the European Regional Development Fund (ERDF) through the NORTE 2030 Regional Programme under Portugal 2030, within the scope of the project BCD.S+M, reference 14436 (NORTE2030-FEDER-00584600), by national funds through FCT - Fundação para a Ciência e a Tecnologia, I.P., under the support UID/50014/2023 (https://doi.org/10.54499/UID/50014/2023), and by FCCN within the scope of the project Deucalion, with reference 2024.00014.TEST.DEUCALION.
This work was also supported by funding from NSF (grants CNS-1956229, CSR-2402328, CAREER-2338457, and CSR-2323100), as well as generous donations from NetApp and Seagate.

\bibliographystyle{IEEEtran}
\bibliography{bibliography}

\end{document}